\documentstyle[prl,epsfig,eqsecnum,aps]{revtex}

\newcommand \bea {\begin{eqnarray} \nonumber }
\newcommand \ee {\end{equation}}
\newcommand \eea {\end{eqnarray}}
\newcommand \be {\begin{equation}}
\newcommand \bi{\bibitem}
\def \l {\langle}
\def \r {\rangle}

\begin{document}
\twocolumn[\hsize\textwidth\columnwidth\hsize\csname@twocolumnfalse\endcsname 

\title{Monte-Carlo simulations
of the violation of the fluctuation-dissipation theorem
in domain growth processes}
\author{A. Barrat}

\address{ International Center for Theoretical Physics\\
Strada Costiera 11,
P.O. Box 563,
34100 Trieste (Italy)}

\date{\today}

\maketitle

\begin{abstract}
Numerical simulations of various domain growth systems are reported, in order
to compute the parameter describing the violation of fluctuation dissipation
theorem (FDT) in aging phenomena. We compute two-times 
correlation and response functions and
find that, as expected from the exact solution
of a certain mean-field model (equivalent to the O(N) model 
in three dimensions, in the limit of N going to infinity),
this parameter is equal to one (no violation of FDT) in the
quasi-equilibrium regime (short separation of times), and zero in the aging
regime. 
\end{abstract}

\vskip.5pc]
\narrowtext

The study of aging phenomena is currently the subject of many efforts, since
this kind of behaviour, for which a given system remains out of
equilibrium at all available times, is present in many systems of
interest, like spin glasses or structural glasses
\cite{review}.
When concerned with the dynamics of a given system, it is
usual to study the correlation function of an observable $A$,
\be
C(t,t')= \l A(t) A(t') \r,
\ee
($\l . \r $ denotes an average over thermal noise)
and the conjugated response function
\be
R(t,t') = \l  \frac{\partial A(t)}{\partial h(t')} \r ,
\ee
where $h$ is an external field applied at time $t'$.
Then, at equilibrium, these two-times quantities satisfy
time translational invariance (TTI: the functions depend only on the
difference of the two times $t-t'$)) and the fluctuation dissipation
theorem (FDT) relating correlation and response by
$R(t-t') = \frac{1}{T} \frac{\partial C(t-t')}{\partial t'}$.
On the other hand, for aging phenomena,
since the dynamics is out of equilibrium, such equilibrium properties are not
expected to hold. In the context of mean-field spin glasses,
Cugliandolo and Kurchan have proposed the general following scenario, in the
limit where the times $t$ and $t'$ go to infinity \cite{cuku}:
for small time differences ($(t-t')/t' << 1$), the system is in 
quasi-equilibrium, and the equilibrium properties hold; 
however, if $t-t'$ is not small with respect to $t'$, the study
of two-time quantities reveals that it is not at equilibrium
($C(t,t')$ depend explicitly on $t$ and $t'$). Moreover, 
they have proposed to measure
the violation of FDT by the function $X(t,t')$ where
\be
R(t,t') = \frac{X(t,t')}{T} \frac{\partial C(t,t')}{\partial t'},
\label{qfdt}
\ee
with the important assumption (afterwards supported by the study of many
different cases, see for example
\cite{cukuledou,franzrieger,franzpspin,yoshino,giorgio}) that, 
as $t$ and $t'$ go to infinity, it becomes a function of times only
through $C(t,t')$:
\be
R(t,t') = \frac{X(C)}{T} \frac{\partial C(t,t')}{\partial t'}.
\ee
This $X(C)$ has moreover received an interpretation in terms
of effective temperature \cite{cukupeliti}.
In the high temperature phase of any system, $X$ is equal to $1$ since the
system equilibrates and the equilibrium properties hold. In
the low temperature phase where aging phenomena appears, violations
of FDT can be quantified by its departure from $1$.
In simulations or experiments, it is more convenient to look at an integrated
response function: the system can be quenched under a magnetic field, 
which is cut off after a waiting time $t_w$ (the relaxation of the
magnetization is then measured, and found to depend on the waiting time), or
it is quenched under zero-field, and a field is applied after $t_w$. In this
second case,
the growth of the zero-field-cooled magnetization
\be
M(t+t_w,t_w) = \int_{t_w}^{t+t_w} R(t+t_w,s) h(s) ds
\ee
is observed.
The quasi-FDT relation (\ref{qfdt}) allows then to write (for a constant
field)
\be
\frac{T}{h} M(t+t_w,t_w)= \int_{t_w}^{t+t_w} X(t+t_w,s) 
\frac{\partial C(t+t_w,s)}{\partial s} ds
\ee
which, in the limit of large $t_w$, gives
\be
\frac{T}{h} M(t+t_w,t_w) = \int_{C(t+t_w,t_w)}^1 X(C) dC .
\ee
Then, if FDT is satisfied, we obtain a linear relation
$\frac{T}{h} M(t+t_w,t_w)= 1- C(t+t_w,t_w)$, independently of
the system, while a deviation from
this straight line in a $M$ versus $C$ plot indicates violation of 
FDT and gives informations on $X$: different systems can have
different types of violation of FDT.
This kind of $M$-versus-$C$ plot has been used to compute the value of $X$
in the aging regime,
analytically for various mean-field models
\cite{cuku,cukuledou,cudean},
and using numerical simulations for the mean-field
Sherrington-Kirkpatrick model \cite{yoshino}, for the $3$-dimensional
Edwards-Anderson model \cite{franzrieger}
(a more realistic spin glass), for the
$p$-spin in finite dimensions \cite{franzpspin}.
While, for the $p=2$ spherical $p$-spin model, equivalent to the
$O(N)$ ferromagnetic model in three dimensions, $X$ is zero
\cite{cudean}, it is found
to be constant for $p \ge 3$, and a non-trivial function of $C$
for the Sherrington-Kirkpatrick and the three-dimensional
Edwards-Anderson model. An numerical investigation of a glass forming
binary mixture (in three dimensions)
has also been made recently \cite{giorgio}, with the result
of a constant value of $X$.

In this letter, we report numerical simulations of various domain-growth
systems (for a review on such systems, see \cite{bray}),
for which it is expected \cite{cukupeliti}
that $X$ is zero in the aging regime.
We examine Ising ferromagnetic systems in two and three dimensions at various
temperatures, and with conserved or non-conserved order parameter.
We also make a simulation of the Edwards-Anderson model in three 
dimensions, to show the striking difference of behaviour.

We consider Ising spins $s_i$ on a square or cubic lattice of linear size
$L$, with ferromagnetic interactions.
Starting from a random configuration, we quench the system at time
$0$ to temperature $T$ and let it evolve according to Glauber dynamics,
with a single-spin-flip algorithm (we will also consider 
later soft-spins evolving through a Langevin equation).
We then measure the spin-spin correlation function
\be
C(t,t')=\frac{1}{N} \sum_{i=1}^N \l s_i(t) s_i(t') \r
\ee
for a unperturbed system. It is known that this correlation function exhibits 
two time regimes: for $t-t' << t'$ (for simplicity we take $t' < t$),
it decays rapidly from $1=C(t',t')$
to $q_{EA}=m^2$, $m$ being the magnetization at temperature $T$; then,
for more separated times, it scales like $L(t)/L(t')$, where $L(t)$ is the
characteristic size of the domains at time $t$.
We also check that the domain sizes
remain much less than $L$, thus ensuring that finite size effects
are not significant. At a certain waiting time $t_w$, we take a copy of the
system, to which a small, constant magnetic field is applied. We then measure
the staggered magnetization 
\be
M(t+t_w,t_w)=\frac{1}{N} \sum_{i=1}^N \overline{ \l s_i(t_w+t) h_i \r} .
\label{magn}
\ee

For spin glasses, the applied field can equivalently be taken uniform
or random, since the interactions between spins are random. 
Taking an uniform field allows to avoid averaging
over the realizations of the field.
On the other hand, for a ferromagnetic system, the action of a uniform
field is to favor one of the phases, which will grow faster. The
correct quantity to measure is therefore the response to a random field:
the staggered magnetization (\ref{magn})
\footnote{in two dimensions, a random field destroys the long range order
(see \cite{nattermann} for a review on the Random Field Ising Model); 
however, the instability destroying it appears only for domain sizes 
growing exponentially with $1/h$ \cite{binder}, so that this effect is not 
important as long as we work with small enough fields
and at times not to long}.
For simplicity, the random $h_i$ are taken from a bimodal distribution
($h_i= \pm h$). The staggered magnetization is
averaged over the realizations of $h_i$, and we checked
linear response using various values of $h$ (typically
from $0.01$ to $0.2$). The sizes used are $L=600$ in two dimensions,
and $L=80$ in three dimensions.

To compare the various curves, obtained for
various systems, temperatures and waiting times $t_w$,
we look at the plots of 
$T M(t+t_w,t_w)/h$ versus $C(t_w+t,t_w)$.
We first made some runs at high $T$: in this case, the system reaches
quickly equilibrium, with TTI ($C(t_w+t,t_w)= C_{eq}(t)$,
$M(t+t_w,t_w)=M_{eq}(t)$) and we checked that FDT holds
($T M_{eq}(T)/h= 1 - C_{eq}(t)$).
For temperatures below the transition temperature, a dependence on $t_w$
appears in $C$ and $M$
(violation of TTI), corresponding to the growth of domains
of the two competing phases. We observe as expected two time regimes
\footnote{we stress that we are interested in long time limits, since the
$X(C)$ function is defined as such; nevertheless, we already can
observe two distinct regimes with finite times, and deduce the
limit of interest.}:
\begin{itemize}
\item for times $t$ smaller than $t_w$, the two-times quantities do not
depend on $t_w$, and FDT also holds:
$T M(t+t_w,t_w)/h = 1 - C(t_w+t,t_w)$. This happens at large values of $C$
(close to $1$) and small values of $M$.
\item for larger times separation, we observe aging in the correlation
function, and also clearly a deviation from FDT.
\end{itemize}

We show the data in figure (1), (2) and (3) for the various systems, and for
various waiting times.
In the aging part, we see that the $M$ versus $C$ curves are in fact
getting  flat, except at small $t_w$. 
A closer look at the data for the aging part
shows that: (i) for larger $t_w$, the plateau reached
by the magnetization is lower, and (ii) for a fixed $t_w$, the magnetization
first grows (like $1-C(t_w+t,t_w)$, this is the non-aging part),
then saturates, and eventually goes slowly
down again, this last effect becoming less important as $t_w$ grows, with a 
flatening of the curves (the slope of this part of the curves 
decreases as $t_w$ increases).
We can explain these effects in the following way:
after $t_w$, the domains have reached a certain typical size, and the domain
walls have a certain total length. The effect of the random field is then to
try to flip some spins; this flipping will be easier at the domain walls, since
the spins there are less constrained by their neighbours.
Therefore we have two
contributions to the staggered magnetization: one from the bulk, and one from
the domain walls. As time evolves, the domains grow and the total
length (or surface, in three dimensions) of the domain walls decreases.
Therefore, the contribution from the interfaces decreases. On the other
hand, the contribution of the bulk will be rather
independent of $t_w$, since the effect on a random field
on a domain of $+$ spins or on a domain of $-$ is the same on average.
The total 
staggered magnetization is thus decreasing when $t_w$ increases, and also,
at $t_w$ fixed, as $t$ grows (after the initial growth, when the field is
switched on). In the limit of large $t_w$, the effect 
of the bulk becomes relatively more important, and we observe the flatening
\footnote{we have checked by a direct visualization of the spins that this is 
indeed what happens: at short times, the majority of the spins flipped by the
random field are on the domain walls, this fraction going then down as the
domains grow; we will also see that this effect due to the motion of domain
walls is not present for the Edwards-Anderson spin-glass.}. 

\begin{figure}
\centerline{\hbox{
\epsfig{figure=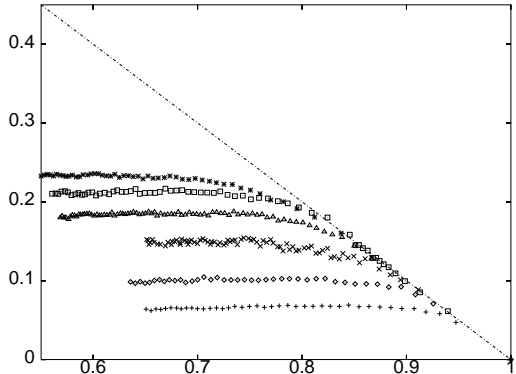,width=5cm,angle=-90}
}}
\caption{$T M(t+t_w,t_w)/h$ versus $C(t+t_w,t_w)$ for two dimensional
domain growth ($T_c=2.27$), at temperatures (from top
to bottom) $T=1.7$ and $t_w=200,\ 400,\ 800,\ 2000$,
$T=1.3$ and $t_w=800$,
$T=1$ and $t_w=800$. The straight line is $M=1-C$: we see that FDT
holds at short times $t$, and the violation of FDT with $X=0$ at longer
time separation.
}
\end{figure}

\paragraph*{Note}:
the reciprocity relations, which state that, for two  observables
$A$ and $B$, the correlations $C_{AB}(t,t')= \langle A(t) B(t') \rangle$
and $C_{BA}(t,t')$ are equal, are also an equilibrium theorem, and 
therefore are not expected to hold for aging dynamics. For
a field $\phi$ evolving according to a Langevin equation, where the
force at time $t$ is $F(t)$, it can be shown \cite{silvio} that, even if
the asymmetry
${\cal A}(t,t')=\langle F(t)\phi(t') - F(t')\phi(t) \rangle $ goes
to zero for long times, the integral $\int_0^t {\cal A}(t,t') dt'$
has a finite limit as $t$ goes to infinity, if the system is
out of equilibrium. Following a suggestion by S. Franz, and
slightly modifying the simulation program, we
checked that this fact, derived using the Langevin equation, also
holds for a Monte-Carlo dynamics, where the field is replaced by
the spins, and the role of the force is 
played by the local field acting on the spins. We therefore
mention this integrated quantity, which
could also be of interest in the studies of aging phenomena.

\begin{figure}
\centerline{\hbox{
\epsfig{figure=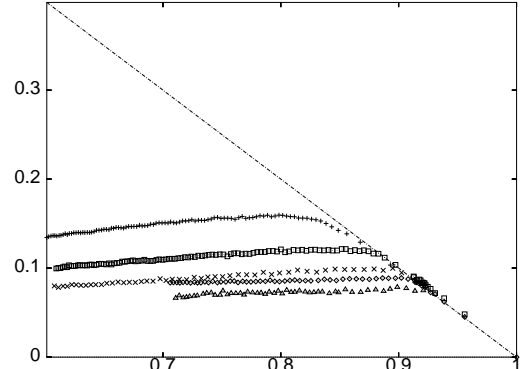,width=5cm,angle=-90}
}}
\caption{Same as figure (1)
for non conserved order parameter
in three dimensions,
$T=2.5$ ($T_c \approx 3.5$), $t_w=100,\ 300,\ 600,\ 1000,\ 1500$.
}
\end{figure}
 
 \begin{figure}
\centerline{\hbox{
\epsfig{figure=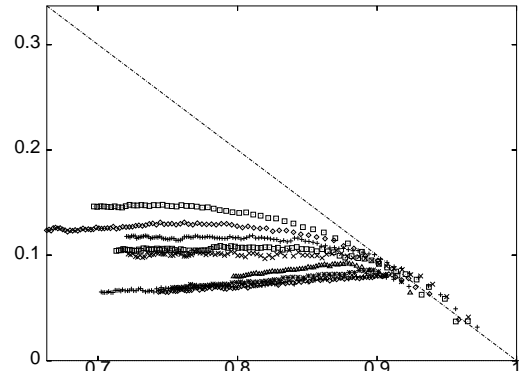,width=5cm,angle=-90}
}}
\caption{Same as figure (1)
for conserved order parameter in two dimensions, $T=0.8$ and
from top  to bottom
$t_w=100,\ 200,\ 400,\ 600,\ 800$, and in three dimensions (lower
symbols),
$T=2$, $t_w=100,\ 200,\ 300,\ 400$.
}
\end{figure}

\paragraph*{Langevin equation}:
since similar results were obtained independently by C. Castellano
and M. Sellitto \cite{claudio} for 
a system of soft-spins evolving through a Langevin equation, we
also mention briefly this case, and show in figure (4) an example 
of the results that can be
obtained with a system of this type: we simulate soft-spins 
on a square lattice, with a quartic potential
confining them to the vicinity of its minima $+1$ and $-1$, and
evolving through the discretized Langevin equation
\begin{eqnarray}\label{lang_discr}
\nonumber
s(i,j,t+1)&=&s(i,j,t)
+ ( s(i+1,j,t) + s(i-1,j,t)  \nonumber\\
&+ & s(i,j+1,t) + s(i,j-1,t)
- 4*s(i,j,t) \nonumber \\
&+& s(i,j,t) - s(i,j,t) ^{3})*h + \eta(i,j,t) \ ,
\end{eqnarray}
where $s(i,j,t)$ is the value of the spin at the lattice site
$(i,j)$ at time $t$, 
$\eta$ is a gaussian noise with zero mean and variance
$2Th$, $h$ being the used time-step. We proceed by parallel
updating of the field, and, at $t=0$, the $s(i,j)$ are taken
as independent random variables uniformly distributed between $-1$ and $1$.
Again, at $t_w$ a random field is switched on and the staggered magnetization
and the correlation are measured.

\begin{figure}
\centerline{\hbox{
\epsfig{figure=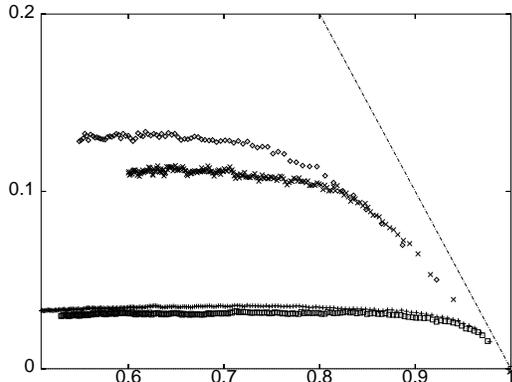,width=5cm,angle=-90}
}}
\caption{Same as figure (1)
for soft spins on a 2-dimensional square lattice,
evolving through (\ref{lang_discr}),
with, from top to bottom, $T=1$ and $t_w=200,\ 600$, $T=0.33$ and
$t_w=200,\ 600$.
}
\end{figure}

All these simulations clearly show that the parameter $X$ is zero
for these domain-growth systems. This flatening of the integrated response
shows that the long-term memory of such systems is in fact weak
\cite{cukupeliti}: the aging
phenomena is essentially in the correlations, while it is also
important for the response in spin-glasses. 

In figure (5), we indeed show the obtained data for an Edwards Anderson
system in three dimensions, with Hamiltonian
\be
H = \sum_{\l ij \r} J_{ij} s_i s_j,
\label{EA}
\ee
where the sum is over nearest neighbours, the spins $s_i$ are Ising spins, and
the couplings $J_{ij}$
are quenched random variables, taking values $+1$ or $-1$
with equal probability.

\begin{figure}
\centerline{\hbox{
\epsfig{figure=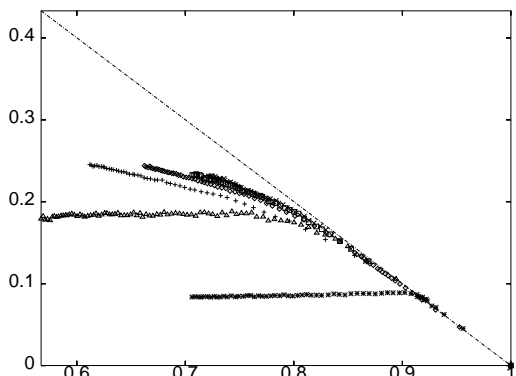,width=5cm,angle=-90}
}}
\caption{Same as in figure (1) for the Edwards Anderson model, $T=0.7$,
from top to bottom $t_w=1000,\ 600,\ 300,\ 100$,
compared to the data for the domain growth in two dimensions
at $T=1.7$, $t_w=1000$ (highest plateau),
and  in three dimensions
at $T=2.5$, , $t_w=1000$ (lowest plateau).}
\end{figure}
We simulated a system of linear size $L=80$ at
$T=0.7$. 
Although no precise conclusion can be drawn as to the form of
the function $X(C)$, since the obtained curves still show a dependence
on $t_w$, it is quite clear (as was shown in \cite{franzrieger})
that they tend to a certain non-trivial curve,
very different from the case of domain growth systems, like the
comparison of figure (5) shows.
Let us remark that curves similar to the ones obtained for the 
EA spin glass have also been obtained for the $p$-spin model in 
three dimensions in \cite{franzpspin} and for the mean-field 
version of (\ref{EA}), the Sherrington-Kirkpatrick model \cite{yoshino}.

To conclude, we have reported measurements of the violation of the
fluctuation dissipation theorem in some systems exhibiting 
domain-growth, and found that, as expected but shown only in one particular
case, the parameter $X$ describing it is equal to zero
in the aging phase (and of course to $1$ in the quasi-equilibrium regime, where
FDT holds).
In the interpretation of \cite{cukupeliti}, this means that the effective
temperature is the temperature of the heatbath in the quasi-equilibrium regime
(corresponding to the fast relaxation of the spins in the bulk of the domains),
while it is infinite in the coarsening regime, which corresponds to the
dynamics of the domains themselves (see \cite{cukupeliti}, paragraph IV-C for a
detailed discussion). 
It should also be noted that this behaviour shows a tendency
of the long-term memory to disappear,
in contrast with spin glasses or glasses.

Acknowledgements: it is a pleasure to thank C. Castellano and S. Franz
for interesting discussions. This work was motivated by discussions
with L. Cugliandolo and J. Kurchan.

\end{document}